\documentclass[twocolumn, a4]{webofc}

\usepackage[varg]{txfonts}
\usepackage{hyperref}
\usepackage{url}
\hypersetup{colorlinks=true,citecolor=blue,urlcolor=blue,linkcolor=blue}

\usepackage{graphicx}
\usepackage{bm}
\usepackage{verbatim}
\usepackage{xcolor}
\usepackage{physics}

\begin{document}

\title{From credible shell model interactions to neutron-capture uncertainties}

\author{
\firstname{Oliver} \lastname{Gorton}\inst{1}\fnsep\thanks{\email{gorton3@llnl.gov}} \and
\firstname{Konstantinos} \lastname{Kravvaris}\inst{1}\fnsep 
}

\institute{Lawrence Livermore National Laboratory}

\abstract{
Nuclear structure theory can provide nuclear astrophysics and nuclear
technologies with bound state properties and transition rates. When describing
nuclear reactions, the list can be extended to include statistical properties
such as nuclear level densities (NLDs) and radiative strength functions (RSFs).
We present the first uncertainty-quantified neutron-capture cross section for
$^{27}$Al based on NLDs and RSFs computed with the shell model (SM). We find
that the USDBUQ500 SM interaction predicts NLDs and RSFs with constant
uncertainties of 6\% and 9\%, respectively. These, in turn, translate to a 5 to
25\% uncertainty in the neutron-capture cross section, which exhibits a
surprisingly non-Gaussian distribution.
}
\maketitle

\section{Introduction} 

To describe neutron-induced reactions in the fast energy region relevant to
nuclear technologies and \textit{r}-process nucleosynthesis (several hundred keV
to several MeV), Hauser-Feshbach (HF) theory is, currently, the tool of choice.
The choice of HF is natural since, after a neutron is absorbed with such
intermediate energies, the decay of the composite system can be described as a
statistical process that depends only on averaged nuclear structure properties.
Those statistical properties, such as nuclear level densities (NLDs) and
radiative strength functions (RSFs), were traditionally modeled using analytic
functions with parameters fit to experimental data. Today, there are numerous
efforts to replace phenomenological models with calculations from nuclear
many-body physics (see~\cite{goriely2018gognyhfb, gorton2026radiative,
zelevinsky2016moments} and references therein). Going further, towards systems
where experimental data is scant, these microscopic models will also be expected
to provide credible uncertainty estimates, a capability that has yet to be fully
realized.

We have recently shown that credible \textit{parametric} uncertainties can be
assigned to phenomenological shell model (SM)
interactions~\cite{gorton2025towards}. The result is an ensemble of interaction
matrix elements which can be sampled to propagate that uncertainty to other
observables. In this work, we sample the USDBUQ500 ensemble of
Ref.~\cite{gorton2025towards} to compute a distribution of neutron-capture cross
sections for aluminum, which is relevant to stellar
astrophysics~\cite{cognata2022exploring}. We produce this distribution of cross
sections by first computing uncertainty-quantified NLDs and RSFs (for a target
nucleus of $^{27}$Al and the compound system $^{28}$Al) as inputs to the HF
reaction code YAHFC~\cite{ormand2021monte}. While not a full account of the
underlying nuclear physics uncertainty, our results represent a significant step
towards understanding the uncertainty budget of nuclear structure inputs to
reaction predictions. We give an overview of the connection between SM
calculations and HF calculations (Sec.~\ref{sec:theory}), present a
proof-of-concept calculation (Sec.~\ref{sec:results}),
and discuss the challenges to be solved going forward (Sec.~\ref{sec:disc}).

\section{Theory}\label{sec:theory}

\subsection{Hauser-Feshbach theory}\label{sec:hf}

The HF cross section can be written as follows:
\begin{equation}\label{eq: hf}
	\sigma_{n,x}(E_n) = 
    \sum_{J^\pi}
	\sigma_{n}^{CN}(E,J^\pi)
    G_{x}^{CN}(E,J^\pi),
\end{equation}
where $\sigma_{n}^{CN}(E,J^\pi)$ is the compound nucleus (CN) formation cross
section; $G_{x}^{CN}(E,J^\pi)$ is the CN decay probability for a given channel
$x = \gamma, n, 2n, ...$; $E$ and $J^\pi$ are the energy and spin-parity of the
CN, while $E_n$ is the energy of the incident neutron. The formation cross
section is determined by an optical model. In this paper we use the global
Koning-Delaroche potential~\cite{koning2003local}, but in a full treatment of
parameter uncertainty, a uncertainty quantified potential like that of
Ref.~\cite{pruitt2023uncertaintyquantified} is required. The CN decay
probability, which is assumed to be independent of the formation event, is given
schematically by:
\begin{equation}\label{eq:G}
    G_{x}^{CN}(E,J^\pi) = 
    \frac{\mathcal{T}^{x}(E, J^\pi)}
    {\sum_{x'} \mathcal{T}^{x'}(E, J^\pi)},
\end{equation}
where for each exit channel $x'$, the generalized transmission coefficient
$\mathcal{T}^{x'}$ is the integral of a particle transmission coefficient
$T^{x'}$ and residual-channel NLD $\rho^{x'}(E_{x'})$:
\begin{equation}
    \mathcal{T}^{x'}(E, J^\pi) = \int T^{x'}(E_{x'})\rho^{x'}(E_{x'})dE_{x'},
\end{equation}
where $E_{x'}$ is the nuclear excitation energy in the channel $x'$. For
brevity, we have omitted details pertaining to sums over individual spins and
orbital angular momentum channels; for a full treatment see the
review~\cite{escher2012compoundnucleara}.

$\mathcal{T}^{x}$ are determined by statistical nuclear structure properties:
the NLDs and particle transmission coefficients, which are required for each
channel entering into the denominator of Eq.~\eqref{eq:G}. For protons and
neutrons, the transmission coefficients are obtained from the same optical model
used to compute the CN formation cross section. The photon transmission
coefficients are computed using RSFs (denoted $f(E_\gamma)$), which are often
assumed to be energy- and spin-independent:
\begin{equation}
    T^{XL}(E_\gamma) = 2\pi E_\gamma^{2L+1} f^{XL}(E_\gamma),\label{eq:tran}
\end{equation} 
where $E_\gamma = E_i - E_f$ is the transition energy between an initial level
$i$ with energy $E_i$ and a lower level $f$ with energy $E_f$. $X$ is the
electromagnetic character of the transition ($X=M$ for magnetic and $X=E$ for
electric) and $L$ is the multipolarity of the transition. 

Various analytic forms for RSFs and NLDs are available in HF codes. Typically,
RSFs are modeled using Lorentzian-based functions~\cite{kopecky1990test,
goriely2019simple}, while NLDs take exponential growth
forms~\cite{gilbert1965composite}. In this work, we begin to replace these
analytic approximations with tabulated calculations from the nuclear shell
model. Due to model space limitations in these calculations we compute only the
M1 RSF, and use the simple modified Lorentzian (SMLO) for the E1 strength with
parameters from Ref.~\cite{capote2009ripl}. For the NLDs we use the brute-force
method. Both approaches are discussed in the next section.

\subsection{Large scale shell model}\label{sec:sm}

The program of the large scale shell model (LSSM) is to (1) obtain an effective
nuclear interaction for a finite valence space, appropriate for the nucleus and
observables of interest, (2) construct a many-body basis from harmonic
oscillator Slater determinants from which the matrix elements of the interaction
can be computed, and (3) compute observables using the interaction (Hamiltonian)
matrix $H$, whether by explicitly converging the real eigenpairs of $H$
(representing the spectra and configuration-space wave functions), or by some
spectral-theory approach (such as the moments method for level
densities~\cite{senkov2010highperformance, ormand2020microscopic,
zelevinsky2016moments}, or strength function methods for electromagnetic
responses~\cite{gorton2026radiative}). In this work, we are concerned with the
impact of the choice of effective interaction matrix elements on the observables
which can be used in reaction calculations. We investigate that impact by
computing NLDs and RSFs for two isotopes, $^{27,28}$Al with the LSSM.

The coupling constants of the effective interaction, which we label $\bm x$,
include the single-particle energies (SPEs) and two-body matrix elements (TBMEs)
of and between the orbitals. The total interaction with one- and two-body terms
can be written as~\cite{brown2006new}:
\begin{equation}\label{eq:int}
    \hat H(\bm x) = \sum_i \epsilon_i \hat n_i
    + \sum_{i\le j, k\le l; JT} V_{ijkl; JT} 
    \hat T_{ijkl; JT}
\end{equation}
where the indices $i, j, k, l$ label the single-particle orbits (typically
harmonic oscillator states); the collective index $i$ is short for all quantum
numbers defining an orbit $(n_i, l_i, j_i)$, with principle quantum numbers
$n_i$, orbital angular momentum $l_i$, and total angular momentum $j_i$. The
number operator for a given shell $i$ is represented by $\hat{n}_i$ and
$\hat{T}$ is the scalar two-body density operator~\cite{brown2006new}.

Historically, the effective matrix elements $\bm x$ are adjusted from their
effective field theory values to minimize the error of predicted binding
energies and low-lying spectra~\cite{brown2006new}. Recently, the fitting
process has been extended to estimate the uncertainty of the fitting
procedure~\cite{fox2020uncertainty, gorton2025towards}. By sampling from the
uncertainty-quantified (UQ) interaction USDBUQ500 from
Ref.~\cite{gorton2025towards}, we can produce UQ NLDs and RSFs, and ultimately,
UQ reaction cross sections.

\subsection{Statistical structure properties from LSSM}\label{sec:nldgsf}

For each of the $N$ realizations of the USDBUQ500 interaction, $\{ \bm x_q
\}_{q=1,...,N}$ ($N=560$), we compute a solution set consisting of the lowest
$M$ nuclear wave functions and corresponding energies: 
\begin{equation}
    \left\{ \psi_i(\bm x_q), E_i(\bm x_q) \right\}_{i=1,...,M}.
\end{equation} 
For each solution set $q$, we produce one NLD and one RSF, yielding $N$
correlated pairs. To compute the NLD for a given set $q$, we explicitly bin the
$M$ energy levels, yielding the shell model NLD:
\begin{equation}
    \rho(E, J, \pi=+) = \sum_i \delta(E' - E_i^{J'}),
\end{equation}
where $\{E_i^J\}_{i=1,...,M}$ is the list of $M$ shell model level energies with
spin $J$. To produce a tabulated result with smooth numerical derivatives, we
either introduce finite-size bins or fold with a Gaussian kernel. In either case
the resolution is chosen similar to the HF energy grid size.

Within the \textit{sd}-valence space used here, we can only compute positive
parity levels. To approximate the full NLD, we assume that the density of
negative parity levels is similar to the positive parity one, but shifted in
energy. We thus estimate the total NLD as:
\begin{align}
    \rho(E) &= \rho(E, \pi = +) + \rho(E, \pi = -) \\
    &\approx \rho(E, \pi = +) + \rho(E-E_{0}^{-}, \pi = +),
\end{align}
where the energy shift $E_0^-$ is taken to be the experimentally known energy of
the first negative parity state. For $^{27}$Al, $E_0^- = 4.055$~MeV and for
$^{28}$Al, $E_0^- = 3.465$~MeV.

To compute the M1 RSF, we use a modified version of the Bartholomew definition
from Ref.~\cite{gorton2026radiative}:
\begin{align}\label{eq:defgsf}
    f^{M1}(E_\gamma) = 
    \left \langle \frac{1}{E_\gamma^{3}} \frac{1}{\Delta E} 
    \sum_{c'} \delta_{j_{c'}j_c} 
    \Gamma^{M1}_{d c'} \right \rangle_{d, j_c},
\end{align}
where $\Gamma^{M1}_{dc}$ are the partial decay widths between the
``compound-nucleus levels'' $c$ and the ``de-excited levels' $d$ so that
$E_\gamma = E_c - E_d$ $\Delta E$ is the energy-averaging bin-width. The sum is
over all $c'$ levels connecting to $d$ which produce a transition with energy
$E_\gamma \pm \Delta E/2$. The average $\langle\cdot\rangle_{j_c}$ together with
the Kronecker-delta yields a spin-averaged strength function, while
$\langle\cdot\rangle_{d}$ averages over all de-excited levels, producing a
function which depends only on $E_\gamma$. The M1 decay widths are computed by
combining standard one-body operators $\mathcal{M}^{\mathrm{M}1}$ (using
coupling constants $g_l^p = 1$, $g_l^n=0$, $g_s=5.5857$, and $g_s^n=-3.8263$)
with all pairs of many-body wave functions with $E_d < E_c$: $\Gamma_{d
c}^{\mathrm{M}1} = 16 \pi k_\gamma^{3}/[9(2j_c+1)] |\langle
\psi_d||\mathcal{M}^{\mathrm{M}1}||\psi_c \rangle|^2$, where
$k_\gamma=E_\gamma/(\hbar c)$.

\section{Results}\label{sec:results}

\begin{figure}[ht!]
    \centering
    \includegraphics[width=.9\linewidth]{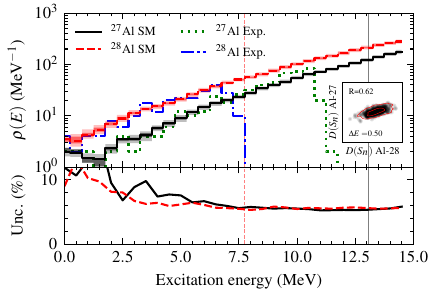}
    \caption{The SM NLDs have a 6\% uncertainty from the underlying interaction.
    There is also a moderate correlation ($R=0.62$, see inset) between the level
    spacings $D(S_n)$ of the two isotopes at their respective neutron separation
    energies (vertical lines). The green dotted and blue dot-dashed histograms
    show the level densities computed from experimentally-known levels.}
    \label{fig:smld}
\end{figure}

By computing the standard deviation of our ensemble of NLD predictions at each
energy, we estimate that the uncertainty generated by the USDBUQ500 interaction
is 6\% (see Fig.~\ref{fig:smld}). The larger uncertainty (bottom panel) at $E<3$
MeV is an artifact of the small number of levels in each bin at lower energies.
The cumulative number of levels as a function of excitation energy has a similar
uncertainty of about 6\%. We also obtain NLD covariance information spanning
both relevant isotopes ($^{27,28}$Al). For the NLD of $^{27}$Al versus $^{28}$Al
at their respective neutron separation energies ($^{27}$Al: 13.058~MeV,
$^{28}$Al: 7.725~MeV), the Pearson correlation coefficient is 0.6
(Fig.~\ref{fig:smld} inset). Correlations between individual excited states (not
shown) reveal that the correlation between levels increases with excitation
energy. This increase is explained by the expectation that highly excited
shell-model levels are strongly mixed with states nearby in energy. 

\begin{figure}[ht!]
    \centering
    \includegraphics[width=.9\linewidth]{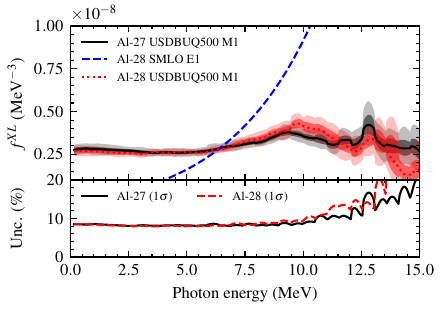}
    \caption{The M1 RSFs computed with USDBUQ500 have a 9\% uncertainty. For
    scale reference, the E1 RSF from the SMLO model used in YAHFC is given.}
    \label{fig:smgsf}
\end{figure}

\begin{figure}[ht!]
    \centering
    \includegraphics[width=.9\linewidth]{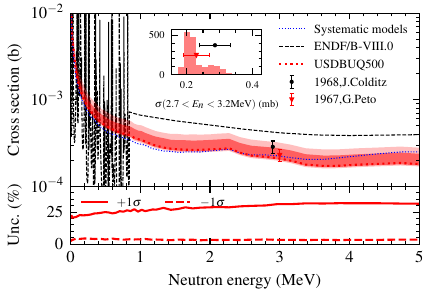}
    \caption{Neutron-capture cross section for $^{27}$Al. The lower panel shows
    the $\pm 1 \sigma$ variance (defined by 16-th and 84-th percentiles). The
    inset shows a histogram of the cross section ensemble between $E_n=$ 2.7~MeV
    and 3.2~MeV, with a markedly non-gaussian. The experimental data from
    \cite{colditz1967measurement} at 2.9~MeV and \cite{peto1967radiative} at
    3.0~MeV are shown in the main panel and the inset.}
    \label{fig:al27cs}
\end{figure}

Similarly, we estimate that the uncertainty in the M1 component of the RSF owing
to interaction matrix elements is 9\% (Fig.~\ref{fig:smgsf}). The apparent
increase in uncertainty above 10~MeV is due to the limited number of individual
transitions used in the calculation. The central line for each nucleus shows the
50-th percentile of predicted RSFs. The dark bands (gray or red) show the
$1\sigma$ predictive interval (PI) approximated by 16-th and 84-th percentiles,
and the light bands show the $2\sigma$ PI approximated by the 1-st and 99-th
percentiles. Along with the E1 RSF from systematic models, the M1 RSF
(Fig.~\ref{fig:smgsf}) were used to calculate the neutron capture cross section
(Fig.~\ref{fig:al27cs}). With this choice of E1 model the M1 RSF is only
relevant to the overall strength below about 10~MeV.

Finally, we show the impact our NLD and RSF uncertainties have on the
neutron-capture reaction on $^{27}$Al (Fig.~\ref{fig:al27cs}). We compare to the
ENDF/B-VIII.0 evaluation~\cite{brown2018endf} and the two most-recent
measurements~\cite{peto1967radiative, colditz1967measurement}. The most striking
feature is that the distribution of cross sections is non-Gaussian and
asymmetric.

\section{Discussion}\label{sec:disc}

With the goal of better equipping astrophysical simulations with robust
estimates of nuclear-reaction uncertainties, the present results illustrate our
vision of the future: by integrating the shell model into statistical reaction
calculations, we move towards a self-consistent and comprehensive description of
cross sections with credible uncertainties. There are, however, limitations to
the present calculations. The main concerns stem from the fact that we must work
in a finite model space. For the NLDs, the SM will always be missing
contributions from configurations outside the model space. This deficiency is
illustrated here by the missing negative-parity states, but there are other
missing configurations such as two-particle-two-hole configurations that break
the $^{16}$O core, for example. For the RSFs, the most obvious omission is the
electric dipole strengths which cannot be modeled in a single-parity model
space. Both of these issues are essentially model defects whose contributions to
the uncertainty is not easily estimated with the parametric uncertainty
techniques we have applied here. Once addressed, uncertainty quantified NLDs and
RSFs from the shell model could be applied to large-scale studies of the impact
of nuclear physics on nucleosynthesis, such as in
Ref.~\cite{kedia2026correlated}.

\section{Acknowledgments}

We thank the RETRO collaboration for useful discussions around this work:
Jeffrey~M.~Berryman, Jonathan~Cabrera~Garcia, Jutta~E.~Escher,
Erika~M.~Holmbeck, Atul~Kedia, Gail~C.~McLaughlin, Cole~D.~Pruitt,
Andre~Sieverding, and Rebecca~Surman.

Prepared by LLNL under Contract DE-AC52-07NA27344 with support from Laboratory
Directed Research and Development project No.~24-ERD-023.

\bibliography{library}

\end{document}